\documentclass{aa}
\usepackage{graphicx}
\usepackage{natbib}
\bibpunct{(}{)}{;}{a}{}{,}

\newcommand\changeddd{}
\newcommand\changedd{}
\newcommand\changed{}

\renewcommand\Re{{\rm Re}}

\newcommand\sgn{\ensuremath {\mathrm{sgn}\,}}
\begin{document}

\title{Drifting Sub-Pulse Analysis Using the Two-Dimensional Fourier Transform}
\titlerunning{Drifting Sub-Pulse Analysis Using the 2DFS}
\author{R. T. Edwards\inst{1} \and B. W. Stappers\inst{1,2}}

\date{}
\institute{Astronomical Institute ``Anton Pannekoek'', 
        University of Amsterdam,
        Kruislaan 403, 1098 SJ Amsterdam, The Netherlands 
  \and
   Stichting ASTRON, Postbus 2, 7990 AA Dwingeloo, The Netherlands}
\offprints{R.~T. Edwards, \email{redwards@astro.uva.nl}}

\abstract{ 
The basic form of drifting sub-pulses is that of a periodicity whose
phase depends (approximately linearly) on both pulse longitude and
pulse number. As such, we argue that the two-dimensional Fourier
transform of the longitude-time data (called the Two-Dimensional
Fluctuation Spectrum; 2DFS) presents an ideal basis for studies of
this phenomenon.  We examine the 2DFS of a pulsar signal synthesized
using the parameters of an empirical model for sub-pulse behaviour.
We show that the transform concentrates the modulation power to a
relatively small area of phase space in the region corresponding to
the characteristic frequency of sub-pulses in longitude and pulse
number. This property enables the detection of the presence and
parameters of drifting sub-pulses with great sensitivity even in data
where the noise level far exceeds the instantaneous flux density of
individual pulses.  The amplitude of drifting sub-pulses is modulated
in time by scintillation and pulse nulling and in longitude by the
rotating viewing geometry (with an envelope similar to that of the
mean pulse profile). In addition, sub-pulse phase as a function of
longitude and pulse number can differ from that of a sinusoid due to
variations in the drift rate (often associated with nulling) and
through the varying rate of traverse of magnetic azimuth afforded by
the sight line.  These deviations from uniform sub-pulse drift
manifest in the 2DFS as broadening of the otherwise delta-function
response of a uniform sinusoid. We show how these phase and amplitude
variations can be extracted from the complex spectrum.
\keywords{Methods: analytical -- pulsars: general} }

\maketitle

\section{Introduction}
Very soon after the discovery of pulsars, it was noticed that the
intensity data sometimes showed ``second periodicities'' within each
pulse, and that the relative phase of this periodic signal drifted in
an organised manner from pulse to pulse \citep{dc68}. The striking
patterns made by drifting sub-pulses in two-dimensional pulse
longitude-time diagrams seemed to be saying something about the
fundamentals of radio pulsar emission, but the question was, and still
is, `what?' Numerous studies of the phenomenon were conducted in the
first few years of pulsar science
(e.g. \citealt{tjh69,sspw70,col70a,bac70b,bac70c,th71,bac73,pag73}),
their frequency steadily decreasing as the limitations of
existing online (recording) and offline (analysis) equipment were
reached. Along with attempts at fitting the data with sub-pulses and
tracking their drift, the Fourier technique developed by Backer in the
references above (now known as the longitude-resolved fluctuation spectrum,
or LRFS) came to be perhaps the most widely used means of examining
the average properties of sub-pulse drift.

With the availability of modern digital back-ends and convenient
offline computing power, we believe that there is now potential for
renewed progress in the field of drifting sub-pulses. We describe in
this paper a means of analysis of drifting sub-pulse data using what
may be considered an extension of the Fourier work of Backer. The
technique was first conceived as a means of detecting and
characterising drifting sub-pulses in the large population of
moderately weak pulsars that have been discovered since the 1970s when
most studies took place. It is also useful, when sufficient
signal-to-noise ratio is available, for the investigation of what
might be considered the fine details of the phenomenon: deviations
from constant amplitude sub-pulses with purely uniform phase
evolution. {\changed We begin with a mathematical description of a
drifting sub-pulse signal as a function of pulse longitude and
pulse number (Sect.~\ref{sec:signal}) and examine its
two-dimensional Fourier response (Sect.~\ref{sec:2dfs}). We then
describe a technique for measuring the deviations of a given signal
from a uniform pure sinusoid (Sect.~\ref{sec:extract}), examine what
is expected under the usual model of drifting sub-pulses 
(Sect.~\ref{sec:carousel}), and use this model as a basis for simulations
testing the applicability of the methods of analysis described here
(Sect.~\ref{sec:sim}).}

\section{Mathematical Model of Drifting Sub-Pulse Signals}
\label{sec:signal}
\subsection{The Form Of Drifting Sub-Pulses}
{\changed The characteristic form of drifting sub-pulses is difficult
to miss in a suitable plot of high-quality data. Plotted as intensity
as a function of pulse longitude and pulse number, a distinctive
``tire tread'' pattern is made as the sub pulses drift monotonically
in longitude with each successive pulse. The form is essentially that
of a windowed two-dimensional sinusoid. The characteristic period of
the sub-pulses in pulse longitude ($P_2$) and pulse number ($\hat{P_3}$
when expressed as a time interval\footnote{We denote the apparent
period $\hat{P_3}$ since it may be an aliased version of the ``true''
$P_3$. See Sect.~\ref{sec:carousel:envelopes}}) may be measured by a
variety of techniques and is usually the first step in any analysis of
pulsars exhibiting drifting sub-pulses.

Naturally the actual signal is not a uniform, pure two-dimensional
sinusoid. The intensity at a given instant depends on the pulse
longitude (the overall emission is, after all pulsed at the spin
period $P_1$) and also has a time variability that may be intrinsic to
the pulsar or induced by interstellar scintillation. The spacing of
the sub-pulses may also vary in a systematic manner with time and/or
pulse longitude. {\changeddd We shall show later (Sect.~\ref{sec:carousel}) that under the most popular model for the origin
of drifting sub-pulses, we may to a very good approximation model these
variations in sub-pulse spacing as as a dependence of $P_2$ on pulse
longitude and/or a time-variability of $P_3$. Observational results
showing that (in specific pulsars) $P_2$ does not vary with time
\citep{pag73,urwe78,la83} and that $P_3$ is independent of pulse
longitude \citep{bac70b,bac70c,kri80,wri81,dls+84,bmhm87} serve to
confirm this assertion. The only reported exception, to our knowledge,
is the sub-pulse modulation reported in the putative pulsar remnant
of Supernova 1987A \citep{mkk+00}, which, unlike other pulsars with
drifting sub-pulses, is probably caused by precession.  We therefore
proceed to describe the signal mathematically on this basis.}

\subsection{Longitudinal Modulation}
The most obvious difference between a true sub-pulse signal and a
purely periodic signal its longitude-dependent amplitude windowing. It
is expected that the sub-pulses will be scaled in amplitude as a
function of pulse phase with a shape similar to the average pulse
profile. We denote this function $a_{\rm l}(\phi)$, where $\phi$ is
the pulse longitude. 

{\changed We may also expect some dependence of longitudinal sub-pulse
spacing on pulse longitude. Expressed another way, the sub-pulses of a
given pulse are periodic in a non-linear function of longitude. The
phase must be at least monontonic and most likely almost linear, or we
would not consider the sub-pulses to be (almost) periodic. We may
therefore treat ``sub-pulse phase'' ($\theta$) in a given pulse as the
sum of a linear term and a longitude-dependent ``phase envelope''
($p_{\rm l}(\phi)$).

The sub-pulse signal for a given pulse can then be represented as}

\begin{eqnarray}
s(\phi) &=& a_{\rm l}A'(\phi)\nonumber\\
& & \sum_{k=1}^{N_h}A_k \sin 
      [k(\phi P_1/ P_2 + p_{\rm l}(\phi) + \theta')+\theta_k],
\end{eqnarray}
\noindent
{\changed where $P_2$ is an ``average'' or ``nominal'' spacing and $A'$
and $\theta'$ are an amplitude scaling and phase shift specific to
this pulse. The model periodic signal (with a period of
$P_2$) } 
is composed of a series of $N_h$ harmonics with amplitude and
phase $A_k$ and $\theta_k$ for each harmonic $k$.  In most instances
the sub-pulses will appear almost sinusoidal and the power of
harmonics with $k>1$ will be small. We therefore neglect the higher
harmonics in further analysis, keeping in mind however that in a real
signal they may be present, with values of $P_2$ and $P_3$ factors of
$k$ smaller than those of fundamental ($k=1$).  Through our choice of
$A'$ and $\theta'$ we may define $A_1=1$ and $\theta_1=0$ so that they
may be neglected in the case of the fundamental.

Since the deviation from pure periodicity is represented by a
longitude-dependent
amplitude scaling and phase shift, we may simply define a complex
modulation envelope $m_{\rm l}(\phi) \equiv a_{\rm
l}(\phi)\exp(ip_{\rm l}(\phi))$, writing the varying sub-pulse
component of a given pulse as 

\begin{equation}
s(\phi) = \Re[m_{\rm l}(\phi)A'e^{i \phi P_1 /P_2 + \theta'}],
\end{equation}
\noindent
where $\Re(x)$ denotes the real part of $x$.  However, it is never the
case that sub-pulses subtract from the overall emission; if the
sub-pulses are to be modelled with sinusoids, we must add a unity bias
which itself is amplitude-modulated by $a_{\rm l}(\phi)$. {\changed In
addition there may be some emission not associated with sub-pulses
(the ``non-drifting'' component, $u(\phi)$), which we include here
under the simplifying assumption that it does not vary from pulse to
pulse. Together these terms are referred to as the ``steady'' part of
the emission\footnote{This is not to say that the steady component is
constant with time, only that it is independent of sub-pulse phase,
that is its form from pulse to pulse varies only by a pulse-specific
amplitude scaling that applies to all longitudes.}. The full signal in
a given pulse can be written:

\begin{equation}
i(\phi) = \Re\left[m_{\rm l}(\phi)A'e^{i \phi P_1/ P_2 + \theta'}\right] 
	+ A'\left[a_{\rm l}(\phi)+u(\phi)\right] .
\label{eq:fullmodel_longitude}
\end{equation}
}

\subsection{Time Modulation}
{\changed In addition to the longitude dependence covered in the
previous section, drifting sub-pulse signals vary on time-scales
longer than one pulse period ($P_1$). We define a time-dependent
amplitude scaling $a_{\rm t}(t)$ which gives rise to the
pulse-specific $A'$ factor used earlier. To describe the
time-dependence of the spacing of sub-pulses in time, we may take a
similar approach and define $p_{\rm t}(t)$ as the difference in phase
between the true signal and a uniform sinusoid of period $\hat{P_3}$.
This function is the source of the pulse-specific $\theta'$ terms of
the previous section.} Again we can define a complex envelope $m_{\rm
t}(t) \equiv a_{\rm t}(t) \exp(ip_{\rm t}(t))$.

\subsection{Full Model of Intensity Signal}
\label{sec:fullmodel}
From the preceding two sections we are able to assemble a model for
the observed intensity of pulsar signals as a function of time and
pulse longitude. {\changed Under the assumption that the 
time-dependent and longitude-dependent modulations are independent,}
we may model the overall modulation as the product of the derived
longitude and time windowing: $m(\phi, t) = m_{\rm l}(\phi)m_{\rm
t}(t)$.  Hence the model of the sub-pulse component  is

\begin{equation}
s(\phi, t) = \Re\left[m(\phi,t)e^{i \phi P_1 /P_2 + 2\pi t/\hat{P_3}}\right],
\label{eq:subpulsesignal}
\end{equation}
\noindent
where %$\theta_{00}$ is the effective sub-pulse phase at $t=\phi=0$, and
$t$ is taken as constant within a given pulse period.  {\changed
We note that in this formulation, if $P_2$ and $\hat{P_3}$ have the
same sign, the sub-pulses appear to arrive earlier with each
successive pulse. We adopt this convention and choose to denote the
opposite sense of drift with a negative value for $\hat{P_3}$.} The
full signal including steady components is:

\begin{eqnarray}
i(\phi, t) &=& a_{\rm t}(t)[u(\phi)+a_{\rm l}(\phi)] \nonumber\\
 &\;&  \mbox{} + \Re\left[m(\phi, t)
                e^{i  (\phi P_1/ P_2 + 2\pi t/\hat{P_3})}\right].
% + \theta_{00})}\right].
\label{eq:fullsignal}
\end{eqnarray}
\noindent

\section{The Two-Dimensional Fluctuation Spectrum}
\label{sec:2dfs}
\subsection{Definition and Properties}
The two-dimensional Fourier transform of a function $f(x, y)$ is
defined as follows:

\begin{eqnarray}
F(u, v) &=& {\mathcal F}[f(x, y)] \nonumber\\
 &=& \int_{-\infty}^{\infty}\int_{-\infty}^{\infty}
   f(x, y)e^{-2\pi i (ux + vy)} {\rm d}x{\rm d}y,
\label{eq:2dfourier}
\end{eqnarray}
\noindent
where ${\mathcal F}$ denotes the Fourier transform operator. Its inverse
differs only in the sign of the exponent:

\begin{eqnarray}
f(x, y) &=& {\mathcal F}^{-1}[F(u, v)] \nonumber\\
 &=& \int_{-\infty}^{\infty}\int_{-\infty}^{\infty}
   F(u, v)e^{2\pi i (ux + vy)} {\rm d}u{\rm d}v.
\label{eq:2dfourier_inv}
\end{eqnarray}

It is usual that the $x$ and $y$ quantities are spatial or temporal,
whilst $u$ and $v$ are spatial or temporal frequencies. The effect of
the two-dimensional Fourier transform is to decompose the input
function into a sum of complex exponentials, or in the case of a
real-valued input function, sinusoids of specific frequency, phase
and amplitude. The basic unit of this decomposition can be seen from
the inverse response to a shifted delta function ($\delta(u-u_0,
v-v_0)$):

\begin{equation}
{\mathcal F}^{-1}[\delta(u-u_0, v-v_0)](x,y) = e^{2\pi i (u_0x + v_0y)}.
\end{equation}
\noindent
The form of the real component of this is simply a sinusoid
in a linear combination of $x$ and $y$, that is a periodic function
with maxima and minima lying on lines of constant $u_0x + v_0y$.

For a discretely, uniformly sampled function with $N_x\times N_y$ samples,
the two-dimensional
Discrete Fourier Transform (DFT)  may be written as
\begin{eqnarray}
F(u, v)  &=& \sum_{j=0}^{N_x-1}\sum_{k=0}^{N_y-1}
   f(j\Delta x, k\Delta y)e^{-2\pi i (uj\Delta x + vk\Delta y)},
\end{eqnarray} 
\noindent
and is computed over a grid of $N_x\times N_y$ values in the
range $-1/2\Delta_x < u < 1/2\Delta_x$, $-1/2\Delta_y < v < 1/2\Delta_y$
where $\Delta_x$ and $\Delta_y$ are the sampling intervals in $x$ and $y$
respectively. As with the one-dimensional DFT, most of the important
theorems concerning the continuous transform also apply to the discrete
case.

When the signal of a pulsar with drifting sub-pulses is displayed as a
function of pulse number and longitude, the patterns made by the
sub-pulses and their drifting are very similar to those of the inverse
Fourier transform of a delta function. This fact alone suggests that
computing the (discrete) Fourier transform of the two dimensional signal
$(i(\phi, t))$ may be an interesting exercise.

Since the signal is real-valued, the result of the Fourier transform 
($I(\nu_l, \nu_t)$) will obey the symmetry 
$I(\nu_l, \nu_t) = I(-\nu_l, -\nu_t)^*$, where the superscript $*$ denotes
the complex conjugate. 
%This is due to the fact
%that we cannot distinguish a negative-positive pair of \nu_l-\nu_t from
%a positive-negative pair, nor a positive-positive pair from a
%negative-negative pair. We can, however, determine whether $\nu_l$
%and $\nu_t$ differ in sign: signals with the same sign in both
%frequencies will have negative gradients in lines connecting
%corresponding sub-pulse peaks, whilst different signs give rise
%to positive gradients. 
From Eqs.~(\ref{eq:subpulsesignal}), (\ref{eq:fullsignal}) and
(\ref{eq:2dfourier}) the result of this operation (which we call the
Two-Dimensional Fluctuation Spectrum or 2DFS) is

\begin{eqnarray}
I(\nu_l, \nu_t) &=&{\mathcal F}[i(\phi, t)]  \nonumber\\
 &=& A_{\rm t}(\nu_t)[U(\nu_l)+A_{\rm l}(\nu_l)] \nonumber\\
 &\;&  \mbox{} + S(\nu_l, \nu_t) \nonumber\\
 &\;&  \mbox{} + S(-\nu_l, -\nu_t)^*,
\end{eqnarray}
\noindent
where functions in upper case
denote the (one- or two-dimensional) Fourier transform of their lower-case
counterparts. %, and we neglect the constant $\theta_{00}$. 
Expanding just
one of the two terms arising from the sub-pulse modulation $s(\phi, t)$,
from the convolution theorem we find

\begin{eqnarray}
S(\nu_l, \nu_t) 
     &=& {\mathcal F}[m_{\rm l}(\phi)m_{\rm t}(t)]\nonumber\\
	& & \mbox{} * \delta[\nu_l-P_1/(2\pi P_2), \nu_t-1/\hat{P_3}] \\
     &=& M(\nu_l, \nu_t) * \delta[\nu_l-P_1/(2\pi P_2), \nu_t-1/\hat{P_3}],
\end{eqnarray}
\noindent
where $*$ denotes the convolution operator. It is also worth noting
that $M(\nu_l, \nu_t)$ is simply equal to
$M_{\rm l}(\nu_l)M_{\rm t}(\nu_t)$ since the non-zero regions of two 
responses lie on orthogonal lines in the two-dimensional spectrum 
and the convolution
reduces to multiplication.

So we see that the 2DFS consists of three main components: one for the
response of the steady part of the pulsar emission, and two
mirror-image components for the sub-pulse modulation. The steady
component is centered at DC ($\nu_l = \nu_t = 0$), whilst the
sub-pulse components are centered approximately at $\nu_l = kP_1/(2\pi
P_2)$ (cycles per radian of pulse longitude), $\nu_t = k/\hat{P_3}$ (cycles
per time unit) where $k=\pm1$.\footnote{\changed If the sub-pulses are significantly
non-sinusoidal, components at other integer values of $k$ will be present.}
Since the modulation envelopes
are not purely real-valued, their Fourier transforms are not expected
to be entirely symmetrical, and there is no unique ``center'' of the
components, however in most cases the phase rotation of the envelope
will be quite minor and a peak around the position specified will be
seen.

The {\changed overall width and shape of the} components depends on
the nature of the modulation envelopes. For a pulsar with no nulling
or scintillation, the steady component will have zero {\changed width}
in the $\nu_t$ direction\footnote{\changed Naturally in practical
analysis of data of limited duration the time windowing leads to an
approximately $\mathrm{sinc}^2$ response on a frequency scale of 1 bin
in the DFT.}, and a one-dimensional inverse Fourier transform of
$I(\nu_l, 0)$ in the $\nu_l$ dimension yields the average pulse
profile. If nulling and/or scintillation are significant, there may be
some {\changed broadening} in the $\nu_t$ direction however very
frequent nulling or scintillation on very short time scales would be
required to {\changed produce a strong, significantly extended component}.
In addition to {\changed the effects of} nulling and scintillation,
the sub-pulse components are broadened due to phase (or $P_3$)
variation over the course of the observation.

{\changed Broadening} of the components in the $\nu_l$ axis derives
from the finite width of the amplitude envelopes ($a_{\rm l}(\phi)$
and $u(\phi)$) and, in the case of the sub-pulse components, due to
phase (or $P_2$) variation across the pulse. 
{\changed The width of this component} is
generally quite large due to the small duty cycle of pulsars, and will
tend to make the spectrum significantly non-zero at $\nu_l = 0$,
$\nu_t = k/\hat{P_3}$ if the number of sub-pulses observable in each
pulse is less than two.

As noted earlier, the frequency extent of the DFT is limited by the
inverse of the sampling interval(s). Co-efficients for frequencies
outside the range may be computed but will always be identical to
those at some (calculable) point in the range due to aliasing. In the
case of the 2DFS, the longitude sampling may be chosen arbitrarily
(within the limits of the hardware and the bandwidth of the received
radiation), but the ``time'' or ``pulse number'' sampling is naturally
set at a value of $P_1$. For this reason, drifting sub-pulse
components outside the range $-0.5 < P_1/P_3 < 0.5$ are
indistinguishable from those within it. 

%his represents no loss of
%information, as shown in
%Sect.\ ~\ref{sec:basic}.  It simply means that all components will appear
%at $\nu_t=1/\hat{P_3}$ rather than the non-aliased $P_3$.

\subsection{Relationship to the Longitude-Resolved Fluctuation Spectrum}
It is readily apparent from Eq.~(\ref{eq:2dfourier}) that the
discrete two-dimensional Fourier transform can be computed by first
computing a one-dimensional DFT along each column of the input data,
and then computing the one-dimensional DFT along each row of the
result (or alternatively, along rows first and columns second). The
result of the first half of this operation when performed on the
pulsar signal $i(\phi, t)$ is known as the longitude-resolved
fluctuation spectrum (LRFS; \citealt{bac70b,bac70c}).

The expected form of this spectrum is easy to understand in the light
of the preceding discussion. Each column (i.e. line of constant
$\phi$) in the spectrum will have three components, as in the
2DFS. The steady component simply appears as a DC term equal to
$u(\phi)+a(\phi)$, whilst the two components for the sub-pulse
modulation appear at $\pm 1/\hat{P_3}$. Calculating the spectrum
of both components (denoted $I_1$ and $I_2$), we find:

\begin{equation}
I_1(\phi, \nu_t) = m_{\rm l}(\phi)e^{i\phi P_1/P_2} \times
	M_{\rm t}(t)*\delta(\nu_t-1/\hat{P_3}) 
\end{equation}
and
\begin{equation}
I_2(\phi, \nu_t) = m_{\rm l}(\phi)e^{-i\phi P_1/P_2} \times
	M_{\rm t}(t)^* *\delta(\nu_t+1/\hat{P_3}) 
\end{equation}
\noindent
That is, the amplitude (or power) spectrum in each phase bin
is identical excluding the scaling factor $a_{\rm l}(\phi)$,
whilst the complex co-efficients are rotated from bin to bin
by an amount $\phi P_1/P_2 + p_{\rm l}(\phi)$.

Due to the presence of a component in both the positive and negative
half of the spectrum, it is impossible to determine the sign of
$\hat{P_3}$ from the {\changed LRF} power spectrum. Signals are effectively
aliased to the range $P_1/\hat{P_3} \in [0,0.5]$.
Only by examining
the sense of phase rotation of the complex co-efficients as a function
of longitude and comparing this to the sign in $\nu_t$ of the component
being considered can the sense of the drift be determined. 

Since the phase relation between longitude bins is expected to be
quasi-periodic, it makes sense to perform a Fourier transform across
each row of the complex spectrum in order to determine the sense of
the drift and its period (i.e. $P_2/P_1$), and to concentrate the
sub-pulse power in a smaller region of phase space for greater
signal-to-noise ratio in $P_3$. The result of this operation is the
two-dimensional DFT of the signal, in other words the 2DFS.

\subsection{Relationship to the Harmonic-Resolved Fluctuation Spectrum}
Recently a new technique for sub-pulse analysis has been devised
by \citet{dr01}. It involves the computation of the so-called
Harmonic-Resolved Fluctuation Spectrum (HRFS). Working from the
original time-series $i(t)$, they compute the one-dimensional spectrum
and ``stack'' it about the $\nu=k/P_1$ harmonics of the signal. Thus,

\begin{equation}
{\mathcal H}[i](x, y) = {\mathcal F}[i]([x+y]/P_1),
\end{equation}
where ${\mathcal H}$ denotes the HRFS and $x$ and $y$ are the
fractional and integer components of the corresponding spectral
frequency multiplied by $P_1$. The HRF power spectrum is usually
plotted with the $x$ parameter as the abscissa (with a domain of 0 to
1) and the $y$ parameter as the ordinate (with integer-valued bins up
to the extent allowed by the sampling interval).

We note that the 2DFS is in fact intimately related (after scaling and
rotation) to the HRFS. The former is formed by stacking the
time-domain data about the pulse period $P_1$ and then Fourier
transforming, whilst the latter is formed by Fourier transforming and
stacking the result about $1/P_1$. That the two should be essentially
identical is not immediately obvious but is easy to show (see appendix
\ref{sec:2dfshrfs}). In fact, the decomposition of the one-dimensional
DFT into a two-dimensional DFT is the basis of parallel FFT codes
\citep{ct65}.

We note that values of $\nu$ giving rise to $0.5 < \nu_t < 1$ are
aliased to $\nu_t-1$ in the 2DFS (i.e. the opposite drift sense),
however this is simply a matter of convention: by the same notion,
signals which truly show the opposite drift sense (that is, negative
$P_3P_2$) are aliased in the HRFS into the region $0.5 < \nu_t
< 1$. In our view, given that there is no reason to prefer one sense
of intrinsic sub-pulse drift over another, specification of
$\hat{P_3}$ in the interval $[-0.5,0.5]$ makes more sense than taking
a preferred drift direction and mapping all drift to this interval
(e.g.[0:1]).

Whilst the two spectra are equivalent, a knowledge of their basis
in the two-dimensional Fourier transform of the longitude-time data
is of value in further analysis, as we shall see below.

\section{Sub-Pulse Analysis with the Two-Dimensional Fluctuation Spectrum}
\label{sec:analysis}
With an awareness of the mathematical basis of the 2DFS in relation
to the form of the drifting sub-pulse signal described in the preceding
sections, a range of analysis strategies can be developed for
the investigation of particular aspects of the sub-pulses. 

The technique was originally conceived with the aim of detecting the
presence and basic parameters ($P_2$ and $P_3$) of sub-pulse drift in
pulsars with limited instantaneous signal-to-noise ratio.  The
sensitivity of this method depends only on the coherence of the drift
and the ratio of its amplitude to that of any non-drifting signal. The
significance of a detection can by improved arbitrarily (for stable
drifters) by simply integrating more pulses, unlike in traditional
single-pulse studies where each individual sub-pulse must be
detectable above the noise. This makes it well-suited to studying the
sub-pulse properties of a large sample of weaker pulsars.

The complex 2DFS is also of use in studying the details of sub-pulse
emission in bright pulsars. In this section we show how the complex
modulation envelopes can be extracted from the spectrum, giving full
information about the deviation from purely periodic sub-pulses.

\subsection{Extraction of the Modulation Envelopes}
\label{sec:extract}
In section \ref{sec:2dfs} we saw that the components arising in the
2DFS due to the sub-pulse drifting take the form of the Fourier
transform of the combined modulation envelope $m(\phi, t)$, shifted to
$\nu_l=P_2^{-1}P_1/2\pi$, $\nu_t=P_3^{-1}$. {\changed To obtain the
envelope from the spectrum, we begin by masking out the steady
component\footnote{This could potentially be achieved by subtraction
of the mean profile from every pulse, however in many cases the power
this removes at non-zero frequency (due to the finite time windowing)
is only a small fraction of that present, due to pulse-to-pulse
intensity variations.} and the mirror-image component. The inverse
transform of this spectrum contains a complex version of the sub-pulse
component of the data which, in the limit of perfect masking of other
components, is its analytic signal.  Its amplitude is only half of
that present in the original data due to the loss of the mirror-image
component, so to circumvent the effect of this in further analysis we
multiply the complex co-efficients of the spectrum by two after
performing the masking. Before performing the inverse transform we
shift the spectrum by the appropriate amount horizontally and
vertically to place the nominal centroid of the sub-pulse response at
DC. The inverse Fourier transform of this shifted spectrum gives the
two-dimensional modulation envelope $\hat{m}(\phi,t)$, as used in
Sect.~\ref{sec:fullmodel}.

In performing the described procedure, is vital that the mirror image
sub-pulse component and the steady component are both carefully
removed from the spectrum. In the case where the conjugate mirror
image components are not separated by regions of nearly zero power, it
is likely that the Fourier transform of the ``true'' complex signal
had power in opposite quadrants, which as a result of the real
sampling is added to that of its conjugated mirror image in the
computed spectrum. Under such circumstances, it is not possible to
reconstruct the true complex signal of the sub-pulse component, and
the derived envelope will differ from that which we had hoped for to a
degree dependent on how significant the overlap is. The same caveat
applies if the drifting component is not clearly separated from the
steady component.

The two dimensional envelope produced using the above procedure can}
be decomposed into longitude- and time-dependent parts under the
assumptions made earlier about the form of drifting sub-pulses, as
such:

\begin{equation}
\hat{m}(\phi,t) = \hat{m_{\rm l}}(\phi)\hat{m_{\rm t}}(t) + {\rm noise}.
\label{eq:decomp}
\end{equation}

We use an iterative scheme to perform this decomposition. By assuming
a form for $m_{\rm l}(\phi)$, one can compute $m_{\rm t}(t)$ through
rearrangement of Eq.~(\ref{eq:decomp}). Using the resultant $m_{\rm
t}(t)$ a new $m_{\rm l}(\phi)$ is computed, and so on. The process is
repeated until convergence, constraining the system to a stable
solution by normalising $m_{\rm t}(t)$ to a mean amplitude of
unity\footnote{There is also a degree of freedom available for arbitrary
phase rotation, however the iterative scheme tends to converge without
the imposition of a constraint to remove this freedom.}.  
Use of simple inversion leads to domination of the inferred phases
by noise, so we use instead the
zero-lag of the normalised cross-correlation function,

\begin{equation}
\hat{m_{\rm t}}(t) = \frac
 {\sum_{j=0}^{N_l-1}\hat{m}(j\Delta\phi,t)\hat{m_{\rm l}}(j\Delta\phi)^*}
 {\sum_{j=0}^{N_l-1}\hat{|m_{\rm l}}(j\Delta\phi)|^2}
\end{equation}
\noindent 
(or the equivalent for $\hat{m_{\rm l}}(\phi)$), which is simply a weighted
version of the longitude-averaged inversion of Eq.~(\ref{eq:decomp}).

{\changed With noiseless data and adequate removal of unwanted
spectral components, a convergent result of this algorithm clearly
gives the correct answer. Under the presence of noise, however, care
must be taken in the interpretation of derived amplitude values.
Denoting the RMS noise in the real and imaginary parts of the
two-dimensional complex envelope {\changedd by} $\sigma_{\rm n}$ (which may be
calculated from the variance of the off-pulse samples in the original
data), the variance of the noise in the real and imaginary parts of
$\hat{m_{\rm l}}$ is
\begin{equation}
\sigma^2_{\rm l} = \frac
 {\sigma^2_{\rm n}}
 {\sum_{j=0}^{N_t-1}\hat{|m_{\rm t}}(jP_1)|^2} .
\end{equation}
\noindent (A similar relation applies for $\sigma^2_{\rm m}$: both
are reduced from $\sigma^2_{\rm n}$ by the integrated power of the
complementary envelope.)

Taking the amplitude envelope incurs some bias due to the presence of
this noise. The derived power values are over-estimated by, on
average, $2\sigma^2_{\rm l}$, leading to significant bias in parts of
the envelope with low (or zero) signal-to-noise ratio. This bias is
intrinsic to the amplitude-phase representation of noisy data, and so
is also familiar from the estimation of the longitude-dependent
envelope from LRF spectra \citep{bac73} and of modulation indices
(e.g. \citealt{th71}). A second-order effect arises from the fact that
the iterative algorithm described above includes the noise bias in its
unity normalization of the time-dependent envelope, leading to less
correlation and hence an unintended attenuation in the next estimate
of the longitude envelope.  The latter effect is easily accounted for
by correction of the normalization to exclude the noise
contribution. Compensating for the former effect requires a more
careful approach.

At the cost of increasing the variance of the noise, the actual
(noisy) values of an envelope may used to determine the bias to
subtract from each element. Alternatively one may take $m^2-2\sigma^2$
as an unbiased estimator of the power envelope, however a problem
remains when amplitudes are to be calculated from negative power
values (such as occurs when the signal-to-noise ratio is
low). Nevertheless, this is the approach that is usually employed in
the calculation of modulation indices.  A final option that may be of
practical use is to make a model longitude-dependent phase envelope
$p_m(\phi)$ (since in most cases the measured phase envelope will be
more smoothly varying than the amplitude envelope). Neglecting errors
in the model envelope, the real part of the product $m_{\rm l}(\phi)
\exp(-ip_m(\phi))$ (corresponding in the complex plane to the
projection of the measured envelope on an axis aligned with the model
phase for that longitude) should be free of bias\footnote{This
estimator also has the virtue of having Gaussian statistics, with
half the variance of that which uses the modulus.}.
}

Once the decomposition is performed, four major classes of questions can
be asked:

\begin{enumerate}
\item How does sub-pulse phase (or equivalently, $P_3$) vary with
time? Are there any systematic variations and if so what causes them?
\item How does sub-pulse amplitude vary with time? 
\item How does sub-pulse phase (or equivalently, $P_2$) vary with pulse longitude? 
\item How does the sub-pulse amplitude window compare to the mean
profile? Is there a non-drifting component?\footnote{If there is no
non-drifting component, the longitude-dependent amplitude envelope
should be identical to the mean pulse profile, see
Eq.~(\ref{eq:fullmodel_longitude}). Any difference can be explained by
the presence of a non-drifting component to the mean profile.}
\end{enumerate}

All of the questions listed above can be (and in some
cases have been) answered to a degree using LRFS, TRFS\footnote{We
call the spectrum formed by performing DFTs along the pulse longitude
axis the Time-Resolved Fluctuation Spectrum (TRFS). It could be used
to examine sub-pulse phase as a function of time, albeit with greatly
reduced sensitivity compared to the techniques presented here, by
examining the phases of the coefficients as a function of time for a
given frequency ($\nu_l$) bin.}  or HRFS by taking one-dimensional
``stacked'' power spectra (for amplitude windowing phenomena) or
one-dimensional ``slices'' (for examining phase relations). However
both these techniques lose sensitivity, in the first case by summing
incoherently and in the second case by using less than the full
available modulation power.  We believe that the technique presented
here is likely to result in measurements of equal or (usually) better
quality since (as long as the assumptions made are valid for the
pulsar) the two envelopes form a complete representation of the
drifting component of the signal.

\subsection{The Carousel Model}
\label{sec:carousel}
\subsubsection{Expected Modulation Envelopes}
\label{sec:carousel:envelopes}
{\changed The most popular conceptual framework for understanding drifting
sub-pulses is that of a carousel-like system of ``sparks'' uniformly
spaced about a ring of constant magnetic elevation and altitude from
the star surface, with the system as a whole slowly rotating about the
magnetic axis \citep{rud72}. As the pulsar rotates and its beam
periodically intersects the observer, a series of pulses, each
composed of a series of sub-pulses is seen.

By treating the problem as if all radiation originates on the surface
of a ``polar cap'', and is beamed in a direction directly opposite to
that of the center of the star (as shown in Fig. \ref{fig:schematic}),
we may consider the rotational geometry to effect, over the course of
each successive pulse, a sampling of the polar cap emissivity along a
locus defined by the intersection of the sight line and the cap.  This
simplification gives a means for good conceptual understanding of the
origin of the observed pulse morphologies and is widely used for this
purpose (e.g.  \citealt{lm88,md99,dr01}), but it must be borne in mind
that in reality the radiation is expected to be beamed along a tangent
to the magnetic field line at the emission location. Therefore the
observed emission at a given instant relates to a plasma flux tube
with its foot at some point on an arc on the polar cap connecting the
depicted ``line of sight'' and the magnetic pole (labelled $\rho$ in
Fig \ref{fig:schematic}c). However, since the field line tangent angle
at a given altitude (where the emission originates) is approximately
proportional to its magnetic colatitude angle, the derived sight-line
locus differs from the true polar cap sampling by a simple scaling of
the magnetic colatitude co-ordinate (see also
\citealt{lm88})\footnote{Furthermore, we note that refraction in the
magnetosphere (e.g. \citealt{pl00}) would also alter the sampling of
the polar cap afforded by the sight line. Again, only the magnetic
colatitude of the derived emission location is corrupted (assuming a
spherically symmetric plasma distribution), but in this case the
mapping from ray angle to emission location is not a simple scaling
and in some cases may not even be one-to-one.}. In any case, the true
sampling effected in magnetic {\it azimuth} is the same as that
derived assuming radial beaming, and if the polar cap emission pattern
is a ring of sparks (or radial ``spokes''), it is this co-ordinate
alone that determines the pulse longitude of the observed sub-pulses.

In systems where the sight-line makes a tangential pass of the ring,
the emission from the sparks is seen as a sequence of one to a few
almost equally-spaced sub-pulses, the phase of which ``drifts'' in
time with respect to a fiducial point in pulse longitude due to the
rotation of the carousel. The number of sparks present, and the
viewing geometry and the angle between the spin and magnetic axes
determine the longitudinal spacing of the sub-pulses ($P_2$), whilst
the time spacing ($P_3$) depends on the rotation rate of the carousel
and the number of sparks it contains.

There are two main effects predicted under the carousel
model that are of importance here. 

Firstly, the carousel rotates continuously (with some period $P_4 =
NP_3$, where there are $N$ sparks), meaning that any variations in
spark intensity, shape or spacing about the ring should manifest as a
periodic feature in the drifting sub-pulse signal. This periodicity
would appear as a periodicity in the time-dependent amplitude and
phase envelopes, giving rise in the 2DFS to a pair of ``sidebands'' at
frequencies $1/P_4$ higher and lower than the primary feature at
$\nu_{\rm t}=1/\hat{P_3}$. This was the startling finding of
\citet{dr01} who applied the HRFS for the first time, to observations
of PSR B0943+10. The detectability of this effect using 2DFS
techniques is discussed further in Sect.~\ref{sec:0943}.

The second effect predicted under the carousel model is that
the longitudinal spacing of the sub-pulses depends on pulse longitude.
The sparks are presumed to be spaced uniformly
in magnetic azimuth ($\psi$) and their separation in spin longitude
is known, but the mapping from pulse longitude to
magnetic azimuth for the sight line is dependent on the degree of
spin-magnetic pole misalignment ($\alpha$) and the viewing geometry
(where $\zeta$ is the angle made between the line of sight and the
positive spin pole).  The relation is:

\begin{equation}
\tan{\psi} = \frac{\sin\phi\sin\zeta}
	{\cos\zeta\sin\alpha - \cos\phi\sin\zeta\cos\alpha} ,
\label{eq:psiphi}
\end{equation}
\noindent where the signs of the numerator and denominator determine
the quadrant for $\psi$ in the usual way (see appendix
\ref{sec:psichi} for derivation), and $\phi$ is measured relative to
the pulse longitude at which the sight-line makes its nearest traverse
of the sight-line.

\begin{figure}
\resizebox{\hsize}{!}{\includegraphics{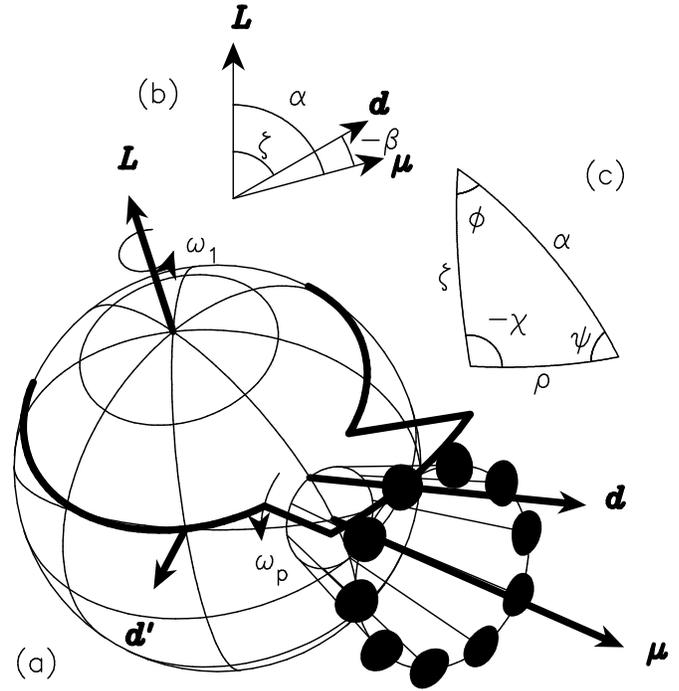}}
\caption{Diagram of sub-pulse emission geometry. In part (a) the
angular momentum, magnetic moment, and line-of-sight vectors are shown
with symbols {\boldmath $L$, $\mu$, and $d$} respectively. The pulsar
rotates about the spin axis with an angular velocity $\omega_1 =
2\pi/P_1$, whilst the ring of sparks that gives rise to the sub-pulses
rotates about the magnetic axis with an angular velocity
$\omega_p=-2\pi /NP_3$, where $N=10$ is the number of sparks.  The
sight-line rotates relative to the pulsar spin frame and is shown here
during the ``on-pulse''. Its intersection with the star surface or the
emitting surface is shown in bold. The co-rotation of the magnetic
field (and the emission associated with the pulse) and the star causes
the sparks to appear as sub-pulses that are amplitude-modulated at the
main pulse (i.e. spin) period. Within this modulation window the
longitudes of sub-pulses drift from one pulse to the next due to their
physical motion about the magnetic pole.  Part (b) shows the main
vectors as they appear in the plane they share when the sight-line
makes its closest approach to the magnetic pole. The angles $\zeta$
(between the sight-line and the spin axis), $\alpha$ (between the
magnetic and spin axes) and $\beta\equiv\zeta-\alpha$ (between the
sight-line and the magnetic axis at their closest approach) are shown.
{\changed In part (a) a second line of sight ($\vec{d'}$) is drawn,
along with the meridian it shares with with the magnetic pole. The
spherical triangle formed by $\vec{L}$, $\vec{\mu}$ and $\vec{d'}$ is
duplicated in part (c). In addition to $\alpha$ and $\zeta$ this
diagram illustrates the meaning of pulse longitude ($\phi$),
polarization position angle ($\chi$, measured relative to the position
angle of the spin axis) and magnetic azimuth ($\psi$) and co-latitude
($\rho$).}  }
\label{fig:schematic}
\end{figure}

The mapping of pulse longitude to magnetic azimuth is monotonic in the
region in which emission is seen, with a gradient of the opposite sign
to $\sgn\beta \equiv \beta/|\beta|$ (where $\beta\equiv\zeta-\alpha$).
In our analysis, $P_2$ and hence {\changedd the rate of change of
sub-pulse phase} 
(${\rm d}\theta/{\rm d}\phi$) are always positive, so
a factor of $-\sgn\beta$ will appear in the expression
{\changedd relating $\theta(\phi)$ to magnetic azimuth ($\psi(\phi)$)}.

The fact that the carousel itself rotates with respect to a fixed
point of magnetic azimuth must also be taken into account. The rotation
period of the carousel is $2\pi/\omega_{\rm p}$ (see fig. \ref{fig:schematic})
where positive $\omega_{\rm p}$ indicates counter-clockwise rotation viewed
from above the active magnetic pole. The direction of drift seen
in the sub-pulses depends also on $\sgn\beta$ : if their signs are
opposite (as is the case in fig. \ref{fig:schematic}) the sub-pulses
will drift toward the trailing edge of the profile. We label this
with a negative value for $P_3$, via 
$P_3 = 2\pi(\sgn\beta)/(N\omega_{\rm p})$. 

Since the sub-pulse phase at a given longitude is only sampled once
per rotation period ($P_1$), we will always observe some periodicity
$-0.5 < P_1/\hat{P_3} < 0.5$. Under the carousel model it is possible
that each spark drifts in longitude by more than half of $P_2$ from
one pulse to the next. In this case we may say that the observed
$\hat{P_3}$ is an ``aliased'' version of the true time taken for the
carousel to drift by $1/N$ turns, $P_3$.  The ambiguity in the
``true'' $P_3$ can be expressed in terms of the (signed, nearest
whole) number of sparks ($n$) that drift by undetected from one pulse
to the next: $P_1/P_3 = n + P_1/\hat{P_3}$.  As shown by \citet{dr01},
this aliasing can potentially be constrained with a measurement of
$P_4$ by observing that $P_4/P_1 = |N/(n+P_1/P_3)|$ and solving for
integer pairs of $n$ and $N$ (in their case $37.35\pm 0.52 =
N/(n-0.4645\pm0.0003)$ implying $n=1$ and
$N=20$)\footnote{\citet{dr01} cited the observed longitudinal
dependence of the polarization position angle as support for this
solution (see Sect.~\ref{sec:pol}).  Although not specifically noted by
\citet{dr01}, the polarization constraint is in fact vital to the
argument: without it other aliasing solutions (all with $N\geq 54$
and $|n|\geq 1$)
are permitted.}

The expression for $\theta(\phi)$ is the sum of three terms, the first
a sign-corrected and scaled version of $\psi(\phi)$
(Eq. \ref{eq:psiphi}), the second a term which accounts for carousel
rotation over the course of the pulse, and the third an offset
representing the carousel position at $\phi=0$.

\begin{eqnarray}
\theta(\phi) &=&
	-N \sgn\beta \tan^{-1} \left[ 
	\frac{\sin\phi\sin\zeta}
	{\cos\zeta\sin\alpha - \cos\phi\sin\zeta\cos\alpha}
	\right] \nonumber \\
&\;&	\mbox{} + \phi \left(n + \frac{P_1}{\hat{P_3}}\right) + \theta'.
\label{eq:thetaphi}
\end{eqnarray}

Near the point of zero longitude, the rate of traverse of magnetic
azimuth is at its maximum:
\begin{equation}
\left|\frac{{\rm d}\psi}{{\rm d}\phi}\right|_{\rm max} = 
\left|-\frac{\sin \zeta}{\sin \beta}\right|.
\end{equation}
\noindent

The apparent rate of traverse of sub-pulse phase is therefore given by
\begin{equation}
\left(\frac{{\rm d}\theta}{{\rm d}\phi}\right)_{\rm max} = N\left|\frac{\sin \zeta}{\sin \beta}\right|
   + n + \frac{P_1}{\hat{P_3}},
\label{eq:maxrate}
\end{equation}
and will generally be close to the ``nominal'' frequency used in 2DFS
analysis (via ${\rm d}\theta/{\rm d}\psi = P_1/P_2$). The difference
between the predicted phase evolution and a constant slope of this
gradient gives the predicted longitude-dependent phase envelope.  An
example is shown in fig. \ref{fig:phasemod}. We note that the
longitude-dependent phase envelope arising from a rotating carousel
will always have a similar form (given sufficient signal to noise
ratio). The gradient of $\theta(\phi)$ is at a maximum at $\phi=0$,
and declines symmetrically on either side.  The spectral shift, if the
correct value of $P_2$ is chosen (and this may be fine-tuned by
examining the resultant envelope), will result in an envelope with
zero gradient at $\phi=0$, and negative gradients at all other
longitudes, with the steepness increasing towards the leading and
trailing edges of the profile.

\begin{figure}
\resizebox{\hsize}{!}{\includegraphics{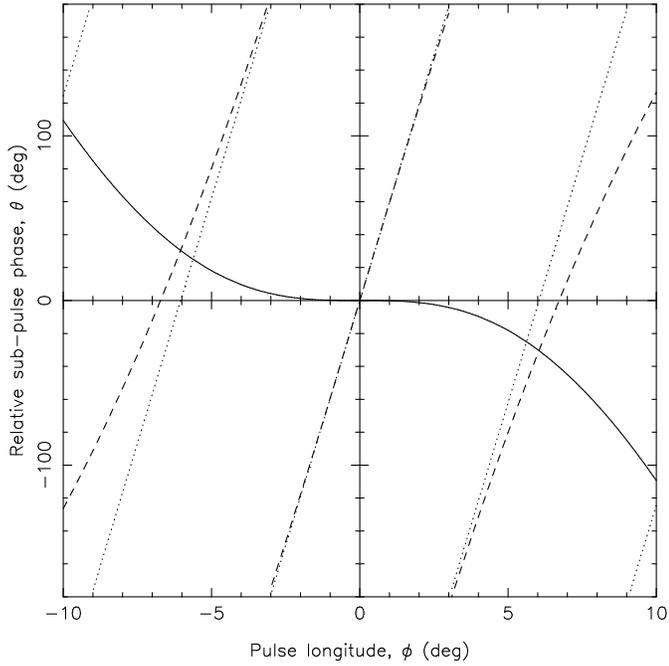}}
\caption{Illustration of the variation in sub-pulse phase ($\theta$)
as a function of pulse longitude ($\phi$). We used $\alpha=10\degr$,
$\beta=2\degr$ and $N=10$ in this example.  The dashed line shows the
value expected through the spherical geometry (Eq.~\ref{eq:thetaphi}),
the dotted line shows that of a pure periodicity with equal phase
slope at $\phi=0$ to the dashed line (Eq.~\ref{eq:maxrate}), and the
solid line shows their difference. This latter quantity can be
considered to be a longitude-dependent phase-modulation envelope to be
applied to the constant periodicity to produce the real signal.}
\label{fig:phasemod}
\end{figure}

The techniques used in this work decompose the sub-pulse signal into
the product of two complex envelopes which depend only on time and
longitude respectively. As such, it is important that the longitudinal
spacing of the sub-pulses does not vary with time, and that the
time-spacing of sub-pulses does not vary with longitude.  The
longitude-independence of $P_3$ is trivially true in the carousel
model, since the carousel rotates as a whole and the angular speed is
constant around the ring. This is seen in LRF spectra of virtually all
pulsars with drifting sub-pulses, see e.g. \citet{bac73}.  The
time-independence of $P_2$ is technically not satisfied under the
carousel model, since the longitudinal spacing depends on $P_3$ (see
Eq. \ref{eq:thetaphi}), which in turn depends on the circulation rate,
which may vary with time. However, for this effect to be significant
Eq.~(\ref{eq:maxrate}) must be dominated by the carousel rotation,
i.e. $|n| \gg 1$. In this case, even minor variations in the rotation
rate could cause the observed (highly aliased) $\hat{P_3}$ to vary
wildly in sign and magnitude. To our knowledge, the only pulsar to
clearly show sense reversals in its drift is PSR B0826$-$34, and the
longitudinal spacing is (within errors) independent of the
instantaneous $P_3$ \citep{bmh+85}. Other pulsars {\changedd that}
show major changes (but not sign reversals) in the drift rate are PSR
B2016+28 \citep{tmh75} and PSR B0809+74 (after nulls;
\citealt{pag73,la83}), both of which also show constant longitudinal
spacing. It therefore appears safe to assume that most systems are not
highly aliased and the longitudinal spacing can be assumed constant
with time.

\subsubsection{Comparison with Polarization Behaviour}
\label{sec:pol}
The longitudinal dependence of the observed position angle of linearly
polarized radiation is expected to be intimately related to carousel
rotation sub-pulse behaviour, and comparisons between polarimetric and
drifting-sub pulse analyses have the potential to provide more
information than either alone. Under the ``magnetic pole'' or
``rotating vector model'' (RVM; \citealt{rc69a}), the position angle
is that of the sky projection of the magnetic field line on which the
emission region at a given instant lies. Analogous spherical geometry
to that used in the derivation of the magnetic azimuth of the sight line
leads to a similar relation:

\begin{equation}
\tan{-\chi} = \frac{\sin\phi\sin\alpha}
	{\cos\alpha\sin\zeta - \cos\phi\sin\alpha\cos\zeta}.
\label{eq:chiphi}
\end{equation}
(See appendix \ref{sec:psichi}. Eq.~(\ref{eq:chiphi}) differs to the
formula of \citealt{kom70} and virtually all later works on pulsar RVM
fitting in that it assumes the standard astronomical convention of
position angle increasing counter-clockwise.)

In principle, $\alpha$ and $\zeta$ can be determined directly by
fitting the observed position angle swing. These values could then be
used with a measurement of the longitude-dependent sub-pulse phase
envelope and Eq.~(\ref{eq:thetaphi}) to determine $N$ and check the
validity of the carousel model (both through its ability to fit the
data and by unambiguous consistency of the fitted value of $N$ with an
integer). In practice there may often be little measurable curvature
in either function, in which case it makes more sense to work with the
(fitted and presumed constant) gradients of the curves. 

The first thing to notice is that the position angle swing may have
either positive or negative slope, depending on the sign of $\beta$.
The maximum rate of position angle swing occurs at the closest point
to the magnetic pole, where ${\rm d}\chi/{\rm d}\phi =
-\sin\alpha/\sin\beta$. If the degree of sub-pulse aliasing is known (or
presumed zero) {\changedd we can} determine the physical direction of
carousel circulation via $\omega_{\rm p} = 2\pi (\sgn\beta)/(NP_3)$.

Secondly, the difference in the absolute value of the slopes can
provide useful information. Since $\sin\zeta/\sin\beta - \sin\alpha/\sin\beta
\simeq \cos (\alpha+\beta/2)$ (for small $\beta$), one may take
\begin{eqnarray}
\frac{1}{N}\frac{{\rm d}\theta}{{\rm d}\phi} -
\left|\frac{{\rm d}\chi}{{\rm d}\phi}\right| 
   &=& 
   \left|\frac{\sin \zeta}{\sin \beta}\right| -
\left|\frac{\sin\alpha}{\sin\beta}\right| 
   + \frac{1}{N}\left(n + \frac{P_1}{\hat{P_3}}\right) \\
&\simeq& \sgn\beta\cos(\alpha+\beta/2) +  \frac{1}{N}\left(n + \frac{P_1}{\hat{P_3}}\right) \label{eq:slopediff}.
\end{eqnarray}
Unless the degree of aliasing is very high (with of the order of one
complete carousel rotation for every star rotation), the second term
may be neglected since it is much smaller in magnitude than the likely
uncertainty in $\cos(\alpha+\beta/2)$. The first term may be estimated
for example from the pulse width (and position angle slope;
\citealt{lm88,ran90}), or taken as unknown to yield $N\simeq ({\rm
d}\theta/{\rm d}\phi) / (|{\rm d}\chi/{\rm d}\phi| \pm 1)$.  In
certain fortuitous circumstances use of this expression may determine
$N$ to better than $\pm 1$, and we could use the nearest integer to
calculate $\alpha$ and $\beta$ using the known position angle and
sub-pulse phase slopes.

We note that the work of \citet{dr01} arrived at ${\rm d}\psi/{\rm
d}\phi-{\rm d}\chi/{\rm d}\phi=-\sgn\beta$\footnote{Deshpande and
Rankin used a different convention, where $\alpha\rightarrow 180\degr
-\alpha$ and $\beta\rightarrow-\beta$.}, seemingly under the
assumption of $\alpha\simeq 180\degr$. Their result of $N=20$ for PSR
B0943+10 remains valid under the analysis of the preceding
paragraph. Taking their value of $34$ for the maximum sub-pulse phase
slope, using the position angle slope of $-2.5$ from \citet{svs88} and
neglecting the (probably significant) uncertainty in this measurement,
we find $N=14 \pm 6$. The marginal consistency may indicate that
$\cos(\alpha+\beta/2)$ is close to $-1$, implying that $\alpha$ is
close to $180\degr$ and the neutron star rotates clockwise as viewed
by the observer, in agreement with \citet{dr01}. If measurements, with
quoted uncertainties, of the position angle and sub-pulse phase slopes
with quoted uncertainties were available, it would be possible to use
the $N=20$ result %with Eq. \ref{eq:slopediff} to quantitatively
to constrain $\alpha$ and $\beta$ without the usual recourse to the use
of pulse width--period relations.  }

\subsection{Tests on Simulated Data}
\label{sec:sim}
Whilst the techniques outlined in Sect.~\ref{sec:extract} are simple and
quite direct, we feel that it is important to demonstrate that they
perform as expected. To this end, we {\changed used the carousel model
to produce several simulated} time series which we subjected to 2DFS
modulation window estimation.

\subsubsection{Form of the Model}
For each sample, we computed the pulse longitude ($\phi$), and the
corresponding magnetic azimuth ($\psi$) that is sampled by the
sight-line at that longitude using Eq.~(\ref{eq:psiphi}) and a
geometry specified by $\alpha$ and $\beta$. This was
converted to a ``sub-pulse phase'' ($\theta$) through multiplication
by $-\sgn\beta N$, and
addition of an offset $\theta_d(t+\phi P_1/2\pi)$ (recalling that $t$
is taken as constant in any given pulse), advancing at a rate of one
turn per interval $P_3(t)$, which produces the drifting as a function
of time. To investigate the detectability of variations in $P_3$, we
simulated the effect of a pulse null as a period of no drifting
($P_3\rightarrow \infty$) during the null, followed by an exponential
recovery to the asymptotic rate $P_3'$. The full form of
$P_3(t)$ is given by the following equation:

\begin{equation}
P_3(t) = \left\{	
              \begin{array}{ll}
                    P_3'& t < t_n\\
                   \infty & t_n < t < t_n+d\\
	            P_3' \left(1-e^{-t/\tau}\right)^{-1} & t > t_n+d
              \end{array}
       \right. 
\end{equation}
\noindent
where $t_n$ is the time at which the null begins, 
$d$ is its duration, and $\tau$ is the time constant
for drift rate recovery. 

For each sample the flux contribution from each spark was computed as
a Gaussian in magnetic azimuth (of width $\sigma_s$ and a peak value
of unity) and the contributions added. The result was scaled by a
factor of $1+c\sin\psi$ ($0 \leq c \leq 1$) to simulate an asymmetry
in the azimuthal flux distribution of the spark carousel, to test for
detectability of the carousel circulation time.

Longitudinal amplitude windowing was applied through multiplication by a
Gaussian in $\phi$ (of width $\sigma_m$, centered
at $\phi=0$ with unity peak), and 
the resulting intensity was in turn scaled by a factor

\begin{equation}
a_t(t) = \left\{	
              \begin{array}{ll}
                    1 & t < t_n\\
                    0 & t_n < t < t_n+d\\
	            1 & t > t_n+d
              \end{array}
       \right. 
\end{equation}
\noindent
to simulate the switching-off of the emission during the null. To this
was added a random noise value (drawn from a Gaussian distribution of
zero mean and standard deviation $\sigma_n$), giving the value
$i(\phi, t)$ to use in later analysis.

{\changed
This model leads to a sequence of pulses, each composed of sub-pulses
whose amplitudes vary (in the specific simulations below) from $\sim
10$\% to $\sim 100\%$ of the unity-peaked overall longitude envelope.
The modulation envelopes measured from the simulated results are
compared with a predicted envelope with amplitudes given by half the
peak-to-peak sub-pulse variation produced by the noiseless model at a
given longitude, after normalizing for a unity-mean time-dependent
amplitude envelope. The predicted longitude envelope peaks at an
amplitude of $\sim 0.45$, compared to $\sim 0.50$ for the average
profile: that is, a small non-drifting component is present, since the
model carousel has some emission at all points on the ring.}

\subsubsection{PSR B0809+74-like Model}
\label{sec:0809fake}
We chose a set of model parameters that give a pulse sequence similar
to those observed from PSR B0809+74 at frequencies around 400
MHz. Specifically, we used a geometry of $\alpha=9\degr$,
$\beta=4.5\degr$ \citep{ran93,ran93b}, requiring (from
Eq.~\ref{eq:maxrate}) $N=16$ to reproduce the reported $P_2/P_1$ of
40~ms/1.29 s $\simeq 32$. A value of $11.0 P_1$ was used for $P_3'$.
The main pulse window had a full width at half-maximum (FWHM) of
$w_{50}=12\degr$ (from $\sigma_m=0.014$ turns), with each sub-pulse
having a FWHM of $5.0\degr$ (from $\sigma_s=0.2/N$). With the
exception of $\alpha$ and $\beta$, all parameters above derive from
\citet{tmh75}.  We included a null at $t_n = 100 P_1$, of duration
$d=8 P_1$ and recovery time constant $\tau=12 P_1$.  No spark-to-spark
intensity variation was included in this model (i.e. $c=0$) since a
time-varying drift-rate complicates the analysis of this
behaviour\footnote{We note, however, that after determining the drift
rate as a function of time, the data could in principle be re-binned
or interpolated to remove the effect of the variation, leaving a pure
periodicity from the circulation.}. We produced two data sets, one
with no noise ($\sigma_n=0$) and one with moderate noise levels
($\sigma_n=0.5$), each with 512 pulses in 4096 longitude bins, of
which only the central 512 were used.

\begin{figure}
\resizebox{\hsize}{!}{\includegraphics{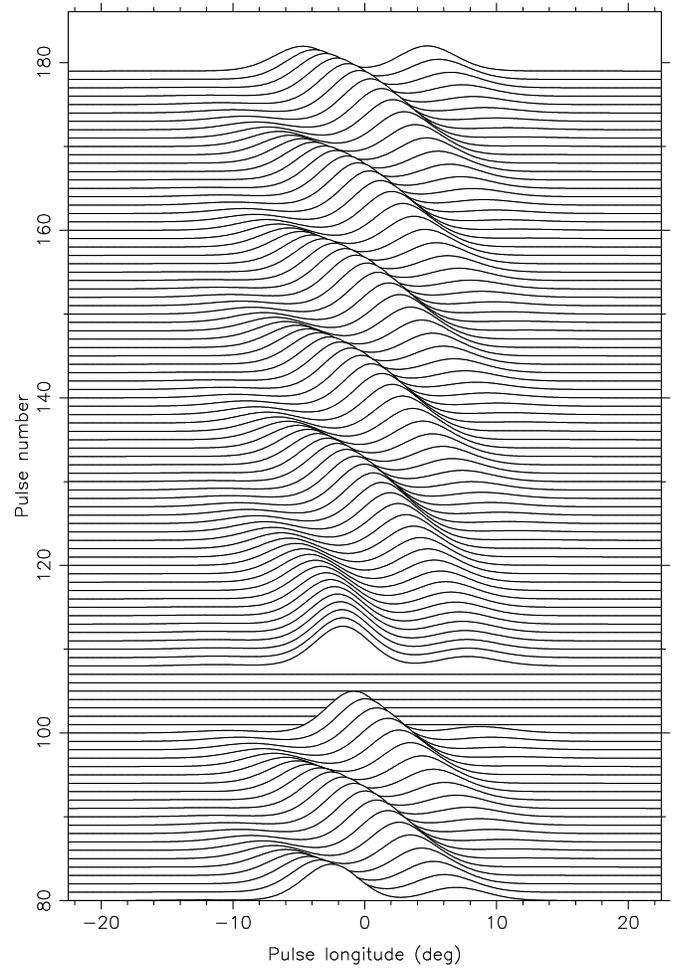}}
\caption{Simulated PSR B0809+74-like data. A sequence of 100 pulses from
the noiseless data are shown, with the interval chosen to show the null.}
\label{fig:0809sim}
\end{figure}

\begin{figure}
\resizebox{\hsize}{!}{\includegraphics{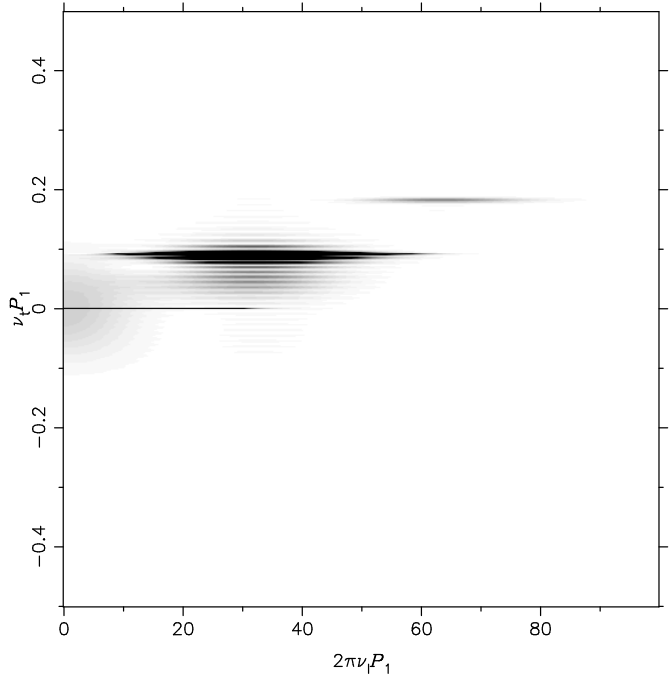}}
\caption{Two-dimensional fluctuation power spectrum of simulated PSR
B0809+74-like data. To enhance low-level structure, grey-levels have
been over-saturated such that pure black corresponds to $10^{-3}$
times the peak power. The resolution used here is higher than that
of the direct DFT and is achieved by zero-padding the input data. }
\label{fig:0809spec}
\end{figure}

The noise-free data and the resultant 2DF power spectrum are shown in
Figs.~\ref{fig:0809sim} and \ref{fig:0809spec} respectively.  The form
of each is as expected from the considerations of the preceding
sections. In particular, the spectrum consists of a DC component, and
the first and second harmonics of the drifting response, with all
components significantly extended in the horizontal axis (due to the
longitudinal windowing) and also at a low level in the vertical axis
(due to the nulling-induced time windowing). {\changed For clearer
presentation, the spectrum in fig. \ref{fig:0809spec} was produced
with enhanced resolution by padding the input data with zeroes before
performing the DFT. In practice this would usually not be done, as it
is unnecessary for the determination of a nominal $\hat{P_2}$ and
$\hat{P_3}$, and complicates evaluation of the significance of
concentrations of power.  We therefore take the position of the
coordinates of the peak bin of the 2DFS computed at the raw
(non-padded) resolution as our nominal periodicity, giving
$\hat{P_2}=P_1/32$ and $\hat{P_3}=P_1/0.092$.  This is in agreement
with the parameters used to generate the data, given the bin size of
$8\times 0.002$ in $P_1/P_2\times P_1/P_3$.}

After forming the full (symmetrical) complex spectrum and shifting it
by the nominal values for $1/\hat{P_2}$ and $1/\hat{P_3}$, the responses
of the average profile and the mirror-image of the fundamental were
removed with a notch filter.  To reduce the degree of ``ringing'' in
the impulse response of the resultant spectrum, the filters
incorporated a gradual transition of the form

\begin{equation}
f(\nu_t) = \left\{	
              \begin{array}{ll}
                    0 &  |\nu_t-\nu_{tc}|/w < 0.5\\
                    2|\nu_t-\nu_{tc}|/w - 1 & 0.5 \leq |\nu_t-\nu_{tc}|/w \leq 1\\
	            1 & |\nu_t-\nu_{tc}|/w > 1
              \end{array}
       \right. ,
\label{eq:filter}
\end{equation}
\noindent
where $f(\nu_t)$ is the filter transmission as a function of frequency
in the $\nu_t$ axis, $\nu_{tc}$ is the centre frequency of the
component to be removed and $w$ is half the total width of the filter,
in this case $w=0.01/P_1$. Since the components to be removed were
broad in the $\nu_l$ axis (and certainly overlapping in this parameter
with the desired component, now shifted to DC), the filter
transmission was independent of frequency in this axis.

The resultant spectrum was then scaled by a factor of two to account
for the use of only one of the two (conjugate-pair) components over
which the modulation power is spread, and the inverse Fourier
transform was taken.  Using the scheme described in
Sect.~\ref{sec:extract}, we then decomposed the resultant two-dimensional
complex modulation envelope into the product of longitude and time
envelopes. The results are shown in Figs.  \ref{fig:0809timeenv} and
\ref{fig:0809longenv}.

\begin{figure}
\resizebox{\hsize}{!}{\includegraphics{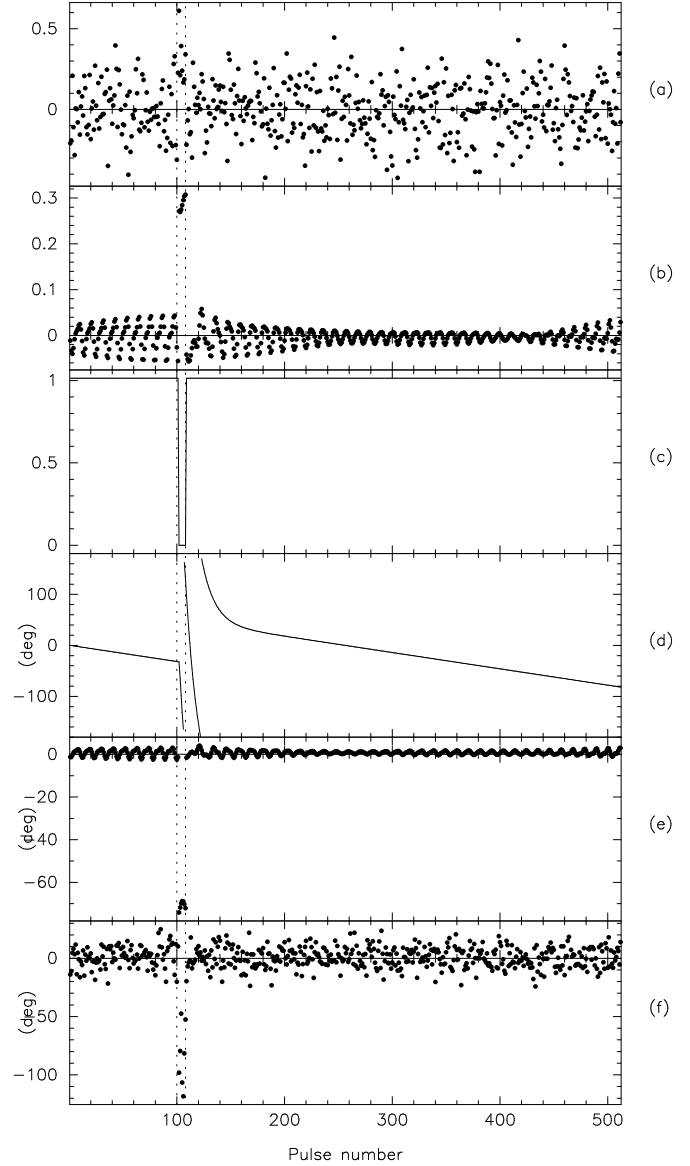}}
\caption{Complex time-dependent modulation envelope and
residuals for PSR B0809+74-like simulated data. The central panels (c
and d) show the amplitude and phase response expected from the
simulation parameters and the chosen nominal $\hat{P_2}$ and $\hat{P_3}$.  
Panels (b) and (e) show the difference between
the measured and expected amplitude and phase envelopes for noiseless
simulated data, whilst panels (a) and (f) show the equivalent
quantities for the noisy data. The dotted vertical lines mark the start
and end of the simulated null.}
\label{fig:0809timeenv}
\end{figure}

\begin{figure}
\resizebox{\hsize}{!}{\includegraphics{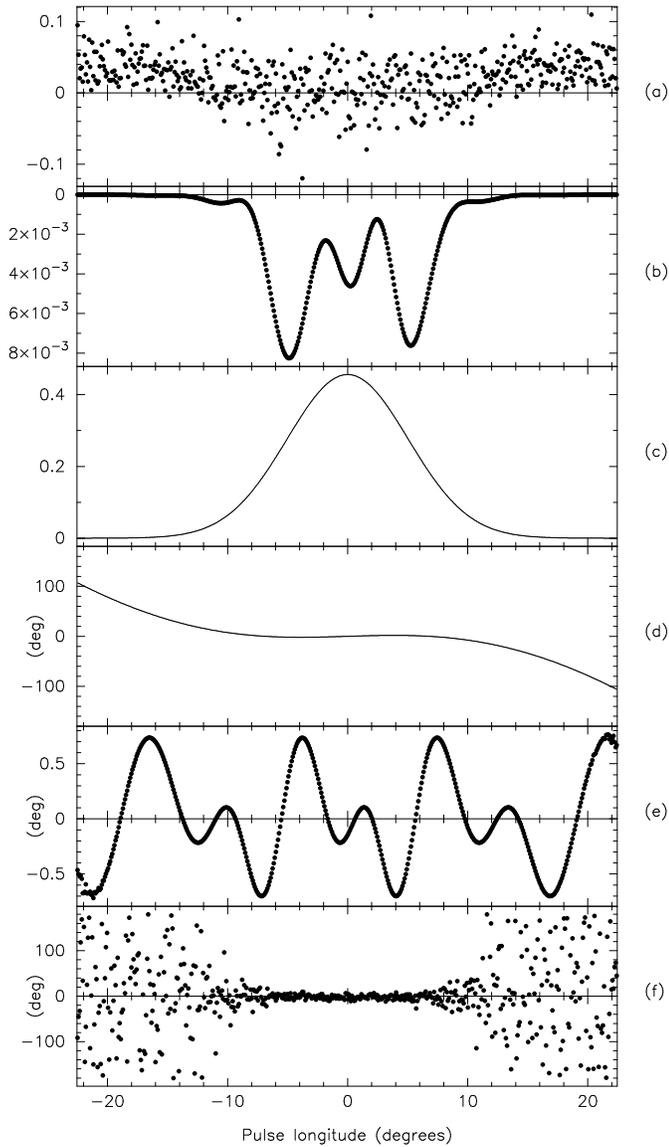}}
\caption{Complex longitude-dependent modulation envelope and residuals for PSR
B0809+74-like simulated data. See caption for
Fig.~\ref{fig:0809timeenv} for descriptions of the data shown in each
panel. 
}  
\label{fig:0809longenv}
\end{figure}

It is evident from the figures that some systematics are present in
the residuals of envelopes of both axes. Beginning with the
time-dependent envelope of the noiseless data, we see that there is
some low-level periodic variation in both the amplitude and phase
terms (panels b and e), with a period equal to $P_3$. We believe that
this is due to a combination of two effects.  The first is the
presence of the second harmonic of the sub-pulse modulation, which is
shifted in the analysis to the former position of the fundamental. The
second is the impulse response of the notch filters used to removed
the DC component. This has a similar form to a sinc function (with a
somewhat shorter time response due to the choice of transfer
function), producing the enhanced oscillations around the null and the
boundaries of the data set. The effect of the latter can be reduced by
making a smoother transition in the filter, however this necessitates
the loss of a greater range of the spectrum, which has the effect of
corrupting both the amplitude and phase response to the post-null
frequency recovery. In our view the phase evolution is more likely to
be of interest than highly accurate amplitude evolution information,
so a narrow filter was chosen. 

As can be seen in panel (e), there is
very little error in the estimated phase, beyond the low-level
zero-mean $P_3$ periodicity.  The residuals of the noisy data appear,
as expected, consistent with the addition of uncorrelated zero-mean
noise to the values estimated from the noiseless data. The degree of
scatter seen here indicates that estimation noise is likely to
dominate over the systematic effects described above for data of all
but the highest of signal-to-noise ratios. An exception is the time
during the null itself, where the phase is of course not measurable,
and the estimated amplitude is affected by the filter response. This
error is trivial to correct by examining the pulse-to-pulse mean flux
density to identify nulls.

The longitude-dependent modulation window (Fig.~\ref{fig:0809longenv})
also shows some systematic error in the estimates made from noiseless
data. We believe this is largely due to errors in the estimated
time-dependent phase envelope discussed earlier. This is a
second-order effect, and its magnitude is very small, in the amplitude
window appearing at the one percent level. The systematic trend in
the amplitude window estimated from the noisy data is a simple
artifact of the representation of noisy complex data as amplitude and
phase values, as discussed in Sect.~\ref{sec:extract}. {\changed
Since the variance
of the noise in the synthesized data was $0.25$, and given the
doubling of spectral coefficients to compensate for filtering out the
mirror image component, the mean power of the noise in the 2D envelope
is expected to be $1.0$ per complex coefficient, giving $\sigma^2_{\rm
n}=0.5$. This leads to a variance of $\sim 1.0\times10^{-3}$ in the
real and imaginary parts of the longitude-dependent envelope (since
$\sum_{j=0}^{N_t-1}|m_{\rm t}(jP_1)|^2 = N_t = 512$), and hence a
bias of up to $\sim 0.045$ at the edges of the envelope. Techniques
for removing this bias are discussed in Sect.~\ref{sec:extract}.  The
second-order effect of underestimation by a normalization error
(Sect.~\ref{sec:extract}) is expected to amount to only $\sim 1$\%,
making it insignificant and not visible in the plots.}

\subsubsection{PSR B0943+10-like Model}
\label{sec:0943}
As an alternative that illustrates the applicability of these techniques
to other observed phenomena, we produced simulated data from a model
with similar parameters to PSR B0943+10. 

We assumed a geometry of $\alpha=11.5\degr$, $\beta=5.4\degr$
\citep{ran93,ran93b} with $N=20$ sparks \citep{dr01}. A profile width
of $20.5\degr$ FWHM \citep{gl98} (giving $\sigma_m=0.023$ turns) was
used, with sparks evenly spaced with widths given by $\sigma_s=0.2/N$,
as in Sect.~\ref{sec:0809fake}.  A value of $P_1/0.5355$ was used for
$P_3'$ \citep{dr01}, with a moderately strong dependence on spark
number ($c=0.2$) applied to give rise to a measurable signal of
carousel circulation. No nulls were included in the simulation, but
the instantaneous $P_3$ values were made to have some variation in
order to model the observed (large but probably finite) $Q$
($=\nu/\Delta\nu$). To this end we produced samples from an
uncorrelated Gaussian noise process and filtered them {\changed in the
frequency domain} with the function $g(\nu)=\exp[-\nu/{0.01 \rm
Hz}]$, to produce a slowly varying function which was scaled to have a
range of $\pm0.05/P_3'$. This was added to to the nominal $1/P_3'$ to
give an instantaneous $1/P_3(t)$. As in Sect.~\ref{sec:0809fake}, one
noiseless and one noisy ($\sigma_n=0.5$) data set was produced. Each
data set contained 512 pulses in 2048 longitude bins, of which only
the central 512 were used.

\begin{figure}
\resizebox{\hsize}{!}{\includegraphics{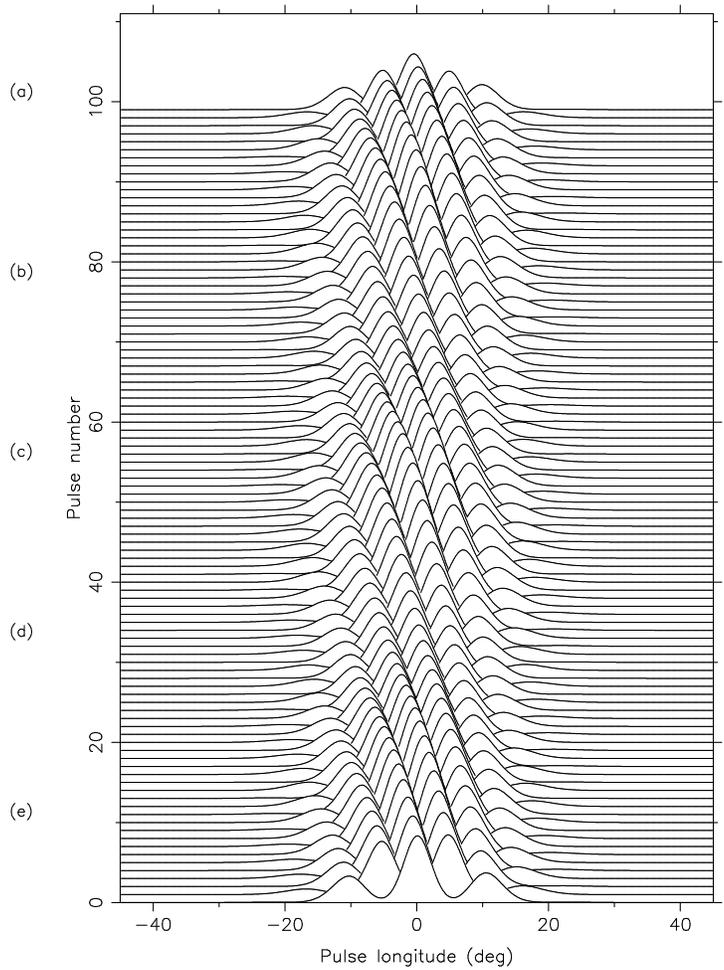}}
\caption{Simulated PSR B0943+10-like data. The first 100 pulses from
the noiseless data are shown. The near odd-even modulation is visible,
as is the amplitude modulation due to the simulated carousel rotation.}
\label{fig:0943sim}
\end{figure}

\begin{figure}
\resizebox{\hsize}{!}{\includegraphics{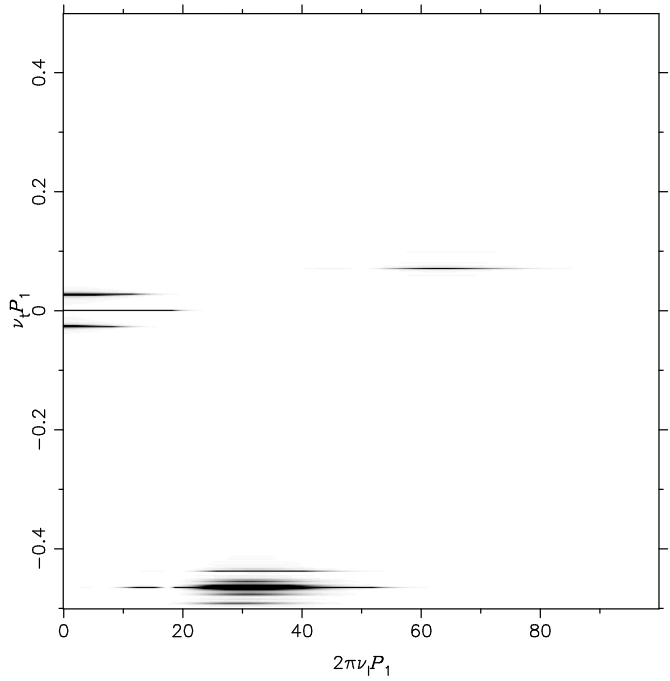}}
\caption{Two-dimensional fluctuation power spectrum of simulated PSR
B0943+10-like data. See caption of fig. \ref{fig:0809spec}.}
\label{fig:0943spec}
\end{figure}

The noiseless data and the resulting 2DF power spectrum are shown in
Figs.~\ref{fig:0943sim} and \ref{fig:0943spec}.  Notice that the
$1/P_3$ feature of the fundamental (at $0.5355 P_1$) is aliased to a
frequency of $-0.4645 P_1$. This reflects the fact that if each
sub-pulse is associated with its closest neighbour in pulse longitude
from the following pulse, its drift appears to progress from earlier
to later longitudes. This approach treats all measured $P_1/\hat{P_3}$
as coming from the range $\pm0.5$.  The conclusion from the work of
\citet{dr01} is that each sub-pulse actually drifts $0.5355$ times the
sub-pulse spacing towards {\it earlier} longitudes with each
successive pulse. Hence, the ``nearest sub-pulse'' ($|P_1/\hat{P_3}| <
0.5$) convention makes a sub-pulse counting error of $n=+1$ (see
Sect.~\ref{sec:carousel:envelopes}). It is important to note that this is
simply a choice of convention. The two cases are indistinguishable
without additional clues as employed by \citet{dr01}.

The 2DFS were filtered with a function of the form shown in
Eq.~(\ref{eq:filter}), with $w=0.03/P_1$, and shifted to place the observed
peak of the fundamental ($P_1/\hat{P3}=-0.465$, $P_1/\hat{P_2}=40$) at
DC. This was scaled by a factor of two and inverse transformed and the
result decomposed into time- and longitude-dependent modulation
envelopes as in Sect.~\ref{sec:extract} and Sect.~\ref{sec:0809fake}. The
results are shown in Figs.~\ref{fig:0943timeenv} and \ref{fig:0943longenv}.

\begin{figure}
\resizebox{\hsize}{!}{\includegraphics{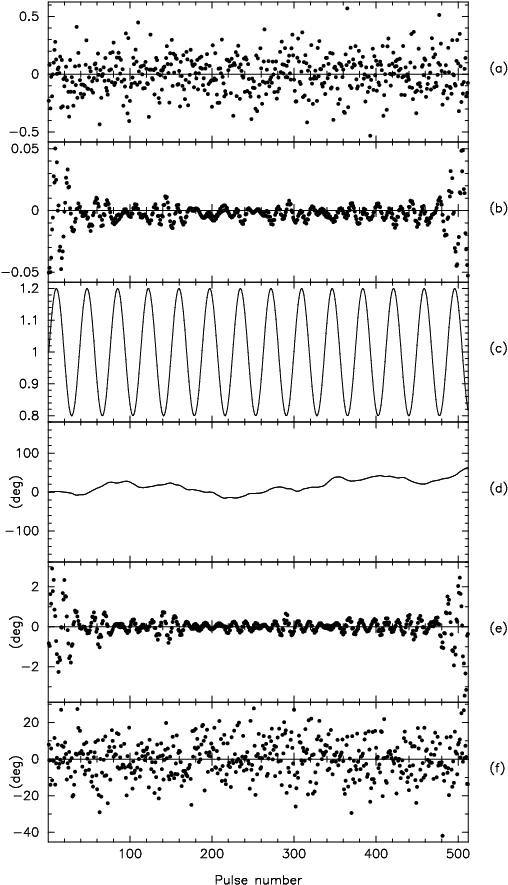}}
\caption{Complex time-dependent modulation envelope and
residuals for PSR B0943+10-like simulated data. See caption for
Fig.~\ref{fig:0809timeenv}.}
\label{fig:0943timeenv}
\end{figure}

\begin{figure}
\resizebox{\hsize}{!}{\includegraphics{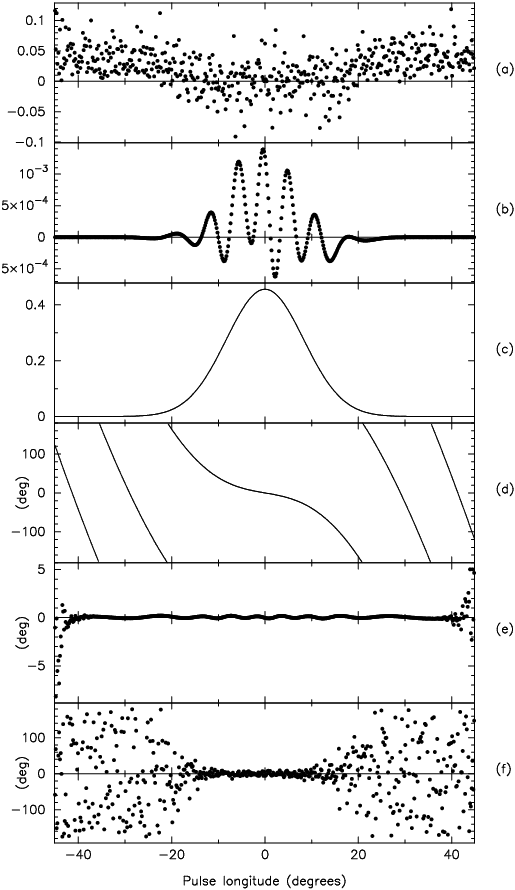}}
\caption{Complex longitude-dependent modulation envelope and residuals
for PSR B0943+10-like simulated data. See caption for
Fig.~\ref{fig:0809timeenv}.}
\label{fig:0943longenv}
\end{figure}

As with the 0809+74-like data, some systematics are present in the
results from the noiseless data. In this case they are less severe due
to the good separation between the fundamental of the drifting
component and the DC component. In any real (noisy) data the errors
are likely to be dominated by noise.  In this model, due to the
broader simulated main pulse, the phase rate (i.e. $P_2$) variation
across the profile is noticeable even in the noisy data. {\changed 
As expected,
the bias in the wings of the longitude-dependent amplitude envelope is
of the same magnitude as that in the 0809+74-like data.}

The amplitude variations due to the simulated carousel circulation are
clearly visible in the modulation envelope estimated from the noisy
data, as are the phase variations due to the simulated frequency
noise. Since the phase variations are quite small, the amplitude
modulation appears highly periodic and can be detected in the power
spectrum of the inferred amplitude envelope.
(Fig.~\ref{fig:0943fake.ampspec}). The modulation also appears as a
pair of ``sidebands'' around the steady and drifting components in the
2DFS (Fig.~\ref{fig:0943spec}). After previous authors
(e.g. \citealt{dr01}), we have ``stacked'' the columns of the power
spectrum corresponding to $1/P_2 < 64$ to produce an ``average''
fluctuation power spectrum. 
Fig.
\ref{fig:0943fake.stackedspec_fund} shows the stacked power spectrum
around the $-0.4645 P_1^{-1}$ drifting component.  The
sidebands associated with the carousel rotation are clearly present,
offset by $0.5355 P_1^{-1}/20\simeq 0.0268 P_1^{-1}$ from their
respective parent features. 

\begin{figure}
\resizebox{\hsize}{!}{\includegraphics{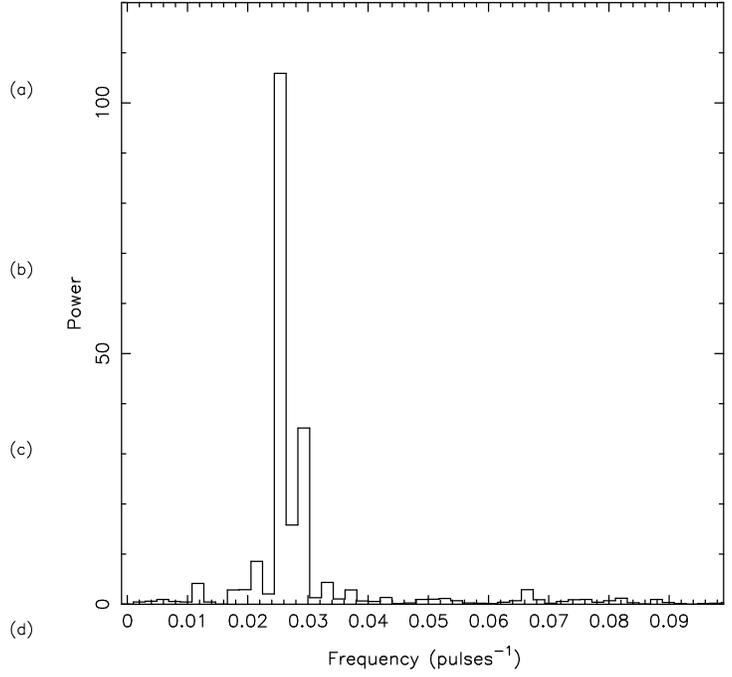}}
\caption{Power spectrum of the time-dependent amplitude envelope
estimated from the noisy data. Values are normalised by the mean of
noise-floor bins.}
\label{fig:0943fake.ampspec}
\end{figure}

\begin{figure}
\resizebox{\hsize}{!}{\includegraphics{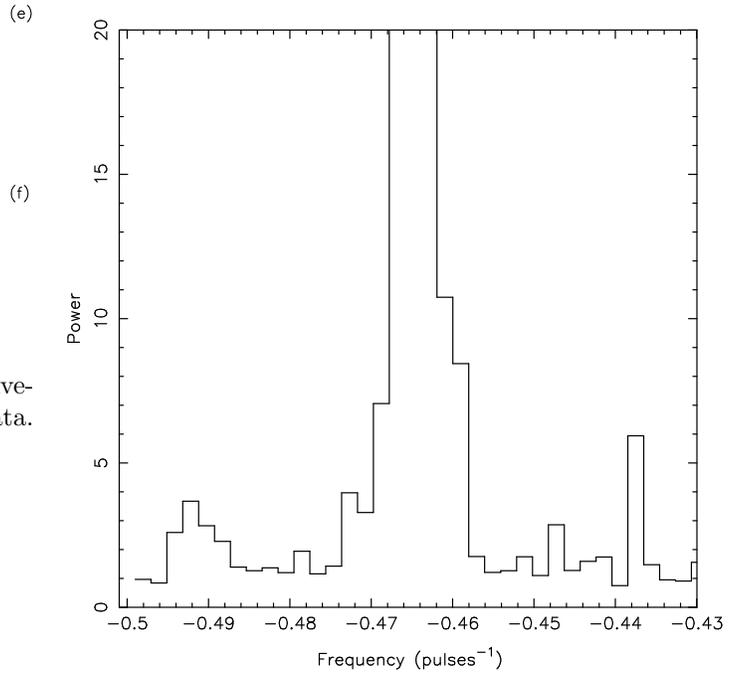}}
\caption{Portion of ``stacked'' 2DFS around the drifting component.
Values are normalised by the mean of noise-floor bins. The central feature
peaks at a value of $\sim487$.}
\label{fig:0943fake.stackedspec_fund}
\end{figure}

The significance of detections of the carousel circulation measured
via the power spectrum of the inferred amplitude envelope versus the
stacked 2DF power spectrum must be evaluated carefully.  Since the
former is estimated from a single Fourier transform, the baseline
noise power {\changed (suitably scaled)} is $\chi^2$ distributed with
two degrees of freedom.  In contrast, the stacked spectrum is formed
by adding 16 columns of the two-dimensional spectrum, so the {\changed
(suitably scaled)} spectral values follow a $\chi^2$ distribution with
32 degrees of freedom.  For a detection of 97.8\% confidence
(corresponding to the 2-sigma point if the noise was Gaussian), the
threshold normalised spectral power values are $\sim 3.8$ for the
power spectrum of the inferred amplitude envelope, versus $\sim 1.6$
for the {\changed normalized} stacked spectrum. The corresponding ``4-sigma''
points are $\sim 10.4$ and $\sim 2.3$. Neglecting effects due to phase
noise, each responds to a modulation of amplitude $c$ with a spectral
power value proportional to $c^2$, plus the ``noise floor'' value of
(on average) unity. From the values of peaks observed in these
simulated data we can infer a proportionality constant of $\sim 20$
between the (noise-subtracted) responses of the two types of
spectra. Hence, values of $c$ giving rise to ``2-'' or ``4-sigma''
detections in the stacked spectrum (with power levels of 1.6 and 2.3)
would produce far more significant detections (with power levels of
$\sim 13$ and $\sim 27$) in the power spectrum of the inferred
amplitude envelope.  This enhancement in sensitivity (amounting to a
factor of $\sim 2$ in the minimum detectable amplitude of carousel
circulation modulation) can be understood as a result of exploiting
the inherent phase relations of the complex spectrum, rather than
(incoherently) summing spectral power values.

As noted earlier, the model assumed here also produces sidebands
around the DC component of the spectrum.  Such sidebands were not
observed in the studies of \citet{dr01}, a fact that needs to be
understood if carousel circulation is to explain the other
sidebands. One possible explanation is that the total (integrated)
luminosity of each spark of the carousel is approximately equal
(leading to no sidebands around DC) but that the width of sparks
differs as a function of position in the carousel.

\section{Conclusions}
We have shown that the two-dimensional Fourier spectrum of pulsar
longitude-time data is of value in the analysis of drifting sub-pulses.
We consider the drifting component of the signal as the convolution
of a complex ``modulation envelope'' with a pure two-dimensional periodicity.
The modulation envelope describes the variations in average sub-pulse
amplitude and phase as a function of longitude and time, and can be
decomposed into the product of two one-dimensional envelopes which
are functions of time and longitude respectively. This makes the
technique well-suited to studying a variety of phenomena associated
with drifting sub-pulses, including longitudinal
variations in sub-pulse spacing (induced due to viewing geometry or
other factors), variations in the drift rate as a function of time
(due to the recovery from a null, random phase noise, etc.),
comparison of the longitudinal amplitude dependence of the drifting and
steady components of the pulsar emission, and variations in the average
sub-pulse amplitude as a function of time (due to nulling, carousel
rotation, etc.). 

\acknowledgements We thank R.~Ramachandran and M.~van der Klis for
helpful comments on the text, {\changed A.~Deshpande for useful
discussion}, {\changedd and the referee (J. Middleditch) for drawing
our attention to the sub-pulse modulation of ``PSR1987A''.}
RTE is supported by a NOVA fellowship.  BWS is supported
by NWO Spinoza grant 08-0 to E.~P.~J.~van den Heuvel.

\appendix
\section{Equivalence of the 2DFS and HRFS}
\label{sec:2dfshrfs}
As noted earlier, the 2DFS and HRFS are in fact equivalent. In this
section we briefly show how this fact arises. 

Consider the response to a single sinusoid, $i(t) = \sin 2\pi\nu t$.
This will appear as a delta function in the HRFS at $x={\rm Frac}[\nu
P_1]$, $y = {\rm Int}[{\nu P_1}]$ (both in units of cycles per
$P_1$-interval), where ${\rm Frac}$ and ${\rm Int}$ denote the
fractional and integer parts of their arguments. For the 2DFS, when
the data are stacked the signal will appear in each pulse period (of
length $P_1$) as a sinusoid with frequency $\nu P_1/2\pi$ (cycles per
radian of pulse longitude) and phase $2\pi t/P_1 \times {\rm Frac}[\nu
P_1]$ where $t/P_1$ is the integer pulse number. Transforming first
across rows {\changed results} in a delta function at
$\nu_l=\nu P_1/2\pi$ with a phase angle of $2\pi t/P_1 \times {\rm
Frac}[\nu P_1]$ (and a corresponding component at $-\nu P_1/2\pi$ which can be
ignored for now as it is simply the source of the symmetry of the
2DFS). Considering the column corresponding to $\nu_l=\nu P_1/2\pi$,
the response is a pure complex exponential with a frequency of ${\rm
Frac}[\nu P_1]/P_1$ (cycles per time unit). Therefore the result of
the two-dimensional DFT is a delta function at $\nu_l=\nu P_1/2\pi$,
$\nu_t={\rm Frac}[\nu P_1]/P_1$.  A 1:1 mapping of $\nu_t=xP_1$,
$\nu_l = y/2\pi$ thus applies in this case, when the fact that the
$\nu_l$ bin size in the discrete 2DFS is $1/2\pi$ is considered. Since
both transformations are linear and 1:1 invertible, we therefore
conclude that they are identical in all cases, given appropriate
mapping between parameters.
{\changed
\section{Derivation of Magnetic Azimuth}
\label{sec:psichi}
The derivation of equation \ref{eq:psiphi} follows a straightforward
coordinate frame transformation, from the spin frame (in which the
sight line parameters $-\phi$ and $\zeta$ represent longitude and
co-latitude), to the magnetic frame. In each frame longitude is
defined such that the positive spin pole of the complementary frame is
at longitude zero.  We definine basis vectors
$[\begin{array}{ccc}\vec{x}&\vec{y}&\vec{z}\end{array}]$ and
$[\begin{array}{ccc}\vec{x'}&\vec{y'}&\vec{z'}\end{array}]$ for the
spin and magnetic frames respectively, aligning $\vec{z}$ (and
$\vec{z'}$) with the positive poles and placing $\vec{x}$ and
$\vec{x'}$ at longitude zero in their respective frames. In what
follows all vectors (including $\vec{d}$ and $\vec{\mu}$) are taken as
being of unit length. The expression for $\vec{d}$ in the spin frame
is:

\begin{equation}
\vec{d} = \left[\begin{array}{c}
              \cos\phi\sin\zeta \\ -\sin\phi\sin\zeta \\ \cos\zeta
	   \end{array}\right]^T
	\left[\begin{array}{c}
		\vec{x} \\ \vec{y} \\ \vec{z}
         \end{array}\right] .
\end{equation}

The transformation between frames may be accomplished by rotating
about $\vec{y}$ by the angle $-\alpha$ (to align the pole with the
magnetic pole), then rotating in longitude (i.e.\ about $\vec{z'}$)
by $\pi$ radians (to place the spin pole at longitude zero).
The transformation is expressed as follows:

\begin{eqnarray}
\vec{d} = & & \left[\begin{array}{c}
              \cos\phi\sin\zeta \\ -\sin\phi\sin\zeta \\ \cos\zeta
	   \end{array}\right]
	\left[\begin{array}{ccc}
		\cos\alpha & 0 & \sin\alpha \\
		0          & 1 &     0       \\
		-\sin\alpha & 1 & \cos\alpha  
	\end{array}\right]
	\left[\begin{array}{ccc}
		-1 & 0 & 0 \\
		 0 & -1 & 0 \\
		 0 & 0 & 1
	\end{array}\right]
	\left[\begin{array}{c}
		\vec{x'} \\ \vec{y'} \\ \vec{z'}
	\end{array}\right] \nonumber \\
 = & & \left[\begin{array}{c}
	 \cos\zeta\sin\alpha - \cos\phi\sin\zeta\cos\alpha \\
	 \sin\phi\sin\zeta  \\
	 \cos\phi\sin\zeta\sin\alpha + \cos\zeta\cos\alpha
	\end{array}\right]^T
	\left[\begin{array}{c}
		\vec{x'} \\ \vec{y'} \\ \vec{z'}
	\end{array}\right] .
\end{eqnarray}

Deriving the longitude of $\vec{d}$ in this frame (i.e. the magnetic
azimuth) via $\tan\psi = d_{y'}/d_{x'}$ yields equation
\ref{eq:psiphi}. This result, after differentiation and correction for
a difference in the sign convention for $\psi$, agrees with that of
\citet{ash88}.

The derivation of the polarization position angle follows a similar
path. It is based in a third frame which is fixed to local rest frame
the observer, has its origin at the star center, positive pole
pointing to the viewer (i.e.\ $\vec{z''} = \vec{d}$) and places the
positive spin pole at longitude zero. All meridians in this frame
project as lines on the sky with position angles offset from the
position angle of the spin axis by their longitude ($\chi$). 
Under the RVM, the position angle of
linear polarization is given by the projection of the local magnetic
field line on the sky, which is given by the projection of the great
circle co-planar with the $\vec{\mu}$ and $\vec{d}$. Hence the
(counter-clockwise) polarization position angle (minus the position
angle of the projected spin axis) is given by the longitude of the
magnetic pole in this frame.  The transformation from the spin frame
is a rotation of $\phi$ in longitude, followed by a rotation about
$\vec{y}$ by $\alpha$ (versus $\zeta$ in the above) and a rotation of
$\pi$ radians about $\vec{y''}$ as above, whilst the input vector has
spherical coordinates $(0,\alpha)$ (versus $(-\phi,\zeta)$ in the
above).  From this it follows that equations \ref{eq:psiphi} and
\ref{eq:chiphi} differ only through
switching of $\alpha$ and $\zeta$ and a sign reversal (in $\phi$
or equivalently, in the final longitude value). 
}

%\bibliographystyle{aa}
%\bibliography{journals_apj,local,modrefs,psrrefs,grbrefs,crossrefs}

\end{document}